\documentclass[preprints,article,accept,moreauthors]{Definitions/mdpi} 

\firstpage{1} 
\pubvolume{1}
\issuenum{1}
\articlenumber{0}
\pubyear{2021}
\copyrightyear{2020}
\datereceived{} 
\dateaccepted{} 
\datepublished{} 
\hreflink{https://doi.org/} 

\Title{Spectral properties of stochastic processes possessing finite
propagation velocity}

\TitleCitation{{Spectral properties of stochastic processes possessing finite propagation velocity}}

\Author{Massimiliano Giona $^{1,}$\orcidA{}, Andrea Cairoli$^{2}$, Davide Cocco$^3$ and Rainer Klages$^4$*}

\AuthorNames{Massimiliano Giona et al.}

\AuthorCitation{Giona, M. ; Cairoli A. ; Cocco D. ; Klages R.}

\address{%
$^{1}$ \quad DICMA,
La Sapienza Universit\`{a} di Roma,
via Eudossiana 18, 00184, Roma, Italy; massimiliano.giona@uniroma1.it \\
$^2$  \quad The Francis Crick Institute, 1 Midland Road, London NW1 1AT, United Kingdom; andrea.cairoli@crick.ac.uk\\
$^{3}$ \quad Dipartimento SBAI,  La Sapienza Universit\`{a} di Roma,
via Antonio Scarpa 16, 00161 Roma, Italy; davide.cocco@uniroma1.it\\
$^{4}$  \quad School of Mathematical Sciences, Queen Mary University of London,
London E1 4NS, United Kingdom; r.klages@qmul.ac.uk }  

\corres{Correspondence: r.klages@qmul.ac.uk} 

\abstract{This article investigates the spectral structure of the
evolution operators associated with the statistical description
of stochastic processes possessing finite propagation velocity.
Generalized Poisson-Kac processes and L\'evy walks are explicitly
considered as paradigmatic examples of regular and anomalous dynamics.
A generic  spectral feature  of these processes
 is the lower-boundedness of the real
part of the eigenvalue spectrum, corresponding
to an upper limit for the spectral dispersion
curve, physically expressing the relaxation rate of a disturbance
as a function of the wave vector.
We analyze also Generalized Poisson-Kac processes possessing
a continuum of stochastic states parametrized with respect to
the velocity.
In this case,
 there exists a critical value of
the wavevector above which the point spectrum ceases to
exist, and the relaxation dynamics becomes controlled by the essential
part of the spectrum. This model can be extended to the
quantum case and, in point of fact, it represents a simple and
 highlighting example
of a sub-quantum dynamics with hidden variables.}


\begin{document}


\section{Introduction}
\label{sec1}

The investigation of micro- and nanoscale physics \cite{KJJ13}, as
well as the extension of statistical physical concepts to new
phenomelogies involving active matter and living beings
\cite{BeDiL16,marchetti2011}, have stimulated the development of more
refined descriptions of stochastic processes, accounting for
background, thermal or quantum fluctuations, aimed at deriving their
statistical features and long-term, in some cases anomalous
\cite{KRS08}, scaling properties \cite{gen1,gen2}.  Experimental
results on the motion of bacteria, amoebae, insects and other classes
of living beings indicate that the classical paradigm of Wiener
fluctuations (the mathematical Brownian motion) does not always apply
to the erratic kinematics of these entities \cite{VLRS11}.

Turning attention to a completely different branch of physics, namely
field theory, the physical constraints on the texture of the
space-time fabric dictate the requirement of finite propagation
velocity as an essential condition for relativistic consistent models
of fluctuations \cite{field1,field2}.  What is remarkable, in this
context, is that the Lorentz covariance applies at the opposite ends
of the length scale spectrum, i.e., at very short (quantum
fluctuations at particle {level}) \cite{quantum} and at very large
lengthscale (cosmological models) \cite{cosmological}.

But even for processes of everyday human experience, such as those
related to mass, momentum and heat transport, representing the natural
realm of investigation of thermodynamic and transport theories, the
quest for overcoming the classical paradoxes resulting from using
stochastic equations with infinite propagation speed, unavoidable
within the assumption of constitutive equations of Fickian type (i.e.,
where the flux of the thermodynamic entities is proportional and
opposite to their concentration gradients, such as in Fick, Fourier
and Newton constitutive equations for mass, heat and momentum
``diffusive'' fluxes), is not only dictated by aesthetical or
epistemological arguments supporting the internal coherence of these
macroscopic and statistical theories \cite{thermo1}, but stems from
the resolution of physical problems such as solute transport through
polymeric materials or heat transport in nanodevices
\cite{polymer,nanoheat}.

A common denominator in all these issues can be found in the
understanding of the emergent properties characterizing stochastic
models possessing finite propagation velocity and of their 
distinctive features, either as regards the local regularity of their
trajectories or their collective statistical behaviour.  
In this article we focus on the latter issue. 
Specifically, {we study} the spectral
structure of the evolution operators 
that propagate the {probability densities of these models}.
Henceforth, we refer to these operators as Statistical Evolution Operators, SEO for
short.  This is because the constraint of bounded velocity enforces
specific and characteristic spectral features, distinguishing them
from their classical and widely used counterparts, namely Wiener and
Wiener-driven fluctuations.

 {We thoroughly investigate fluctuation spectra of} 
 Generalized Poisson-Kac processes (GPK)
 \cite{kac,pinsky,kolesnik,gpk1} and L\'evy Walks (LW)
 \cite{lw1,lw2,lw2a,lw3},    
 {as these processes represent two of the main classes of stochastic models} 
 characterized by bounded propagation velocity. 

{GPK processes originate from the need of defining a  sufficiently wide class
of stochastic models able to interpret at a mesoscopic level the equations of extended thermodynamics
for physical processes far from equilibrium \cite{thermo1,jou}. This represents an extension
of the original model proposed by Kac \cite{kac} of a stochastic process driven 
by the parity of a Poisson counting process, which have been subject of intense investigation
in the past as a prototype of a non-Markovian stochastic motion driven by bounded, dichotomous and colored
noise \cite{hanggi,weiss,horst,bena,onofrio}.}
 {We choose} these two classes of processes 
 {because} they are subjected to a common and complete
 statistical description in terms of the partial probability densities
 parametrized with respect to internal variables of the process
 {(for GPK, the velocity direction; for LWs, the transition age)}. 
 {This was shown especially for LWs} 
 by the recent work of Fedotov \cite{fedotov}
 (see also \cite{fe1,fe2,fe3,jphysa}), and by the unifying theory of
 Extended Poisson-Kac processes developed by Giona and collaborators \cite{epk}.
 {Our main result is} that the real part of the
 eigenvalues of the associated SEO {of these processes} is bounded from below. 
 Consequently, any perturbation of arbitrary wavelength cannot relax
 faster than an exponential function $e^{-\mu_l t}$, 
 {where $\mu_l$ is such lower bound.}
 
The article is organized as follows. 
{In Sections \ref{sec2} and \ref{sec3}, we discuss one-velocity models. 
These are processes characterized by a single velocity in norm, 
where only the direction of motion may change.} 
These processes may seem, at a first glance, a simplistic physical model. 
In point of fact, they represent an important physical situation related
to photon propagation in material media. 
{In this scenario, in fact,} the transition in
the velocity direction {of the photon is solely} dictated by scattering events \cite{photon1}.

{In Section~\ref{sec4},} we extend the analysis to
processes characterized by  a continuous structure of
stochastic states. While the main qualitative result
obtained in Sections \ref{sec2} and \ref{sec3}
{is confirmed} also in this case, 
novel and qualitatively interesting features arise,
such as the incompleteness of the spectrum, and the fact that the
eigenvalue spectrum can be defined solely for particular wavevectors
(in the case discussed, these belong to an interval). 
The consequence of the
latter property
is that   the relaxation decay in
density dynamics is controlled not only by the point
spectrum but also by the essential  component of the spectrum.
For a mathematical definition of the point and essential spectra the
reader is referred to the classical monograph by Kato \cite{kato}.
Throughout this article we adopt an ``operational'' interpretation of the
essential spectrum as the spectral complement to the point spectrum:
an eigenvalue of the linear operator $A$ defined in a functional space $F$
(say the space of square summable functions) belongs to the essential
spectrum if the corresponding eigenfunction does not belong to
the functional space $F$.

{Given the strong analogy between Poisson-Kac processes
  and quantum mechanics, provided by the seminal paper by Gaveau et
  al. \cite{gaveau}, we explore an extension of the Gaveau model based
  on the functional structure of the model {developed in
    Section~\ref{sec4}} possessing a continuum of velocity states,
  focusing on its spectral properties.  This provides an interesting
  and exactly solvable example of a prototypical sub-quantum theory,
  in the meaning of {David Bohm \cite{bohm1,bohm2}, the} emergent
  properties of whose theory coincide with the classical Schr\"odinger
  equation.}

\section{One-velocity GPK processes}
\label{sec2}

\begin{figure}
\begin{center}
\includegraphics[width=12cm]{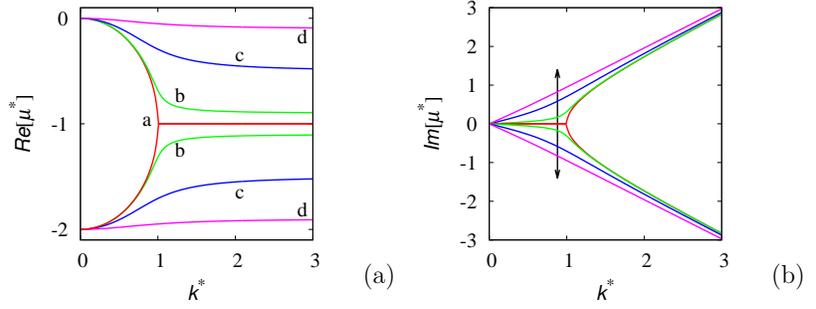}
\end{center}
\caption{Real (\textbf{a}) and imaginary  (\textbf{b}) part of
the eigenvalue spectrum $\mu^*$ vs $k^*$ of the 1d-biased GPK model
at different values of the parameter $r$. Panel (a): line
(a) refers to $r=0$, line (b) to $r=0.1$, line (c) to
$r=0.5$, line (d) to $r=0.9$. Panel (b): the arrow
indicates increasing values of $r$, the same as in panel (a).}
\label{Fig1}
\end{figure}

Consider  the simplest GPK process
on the real line possessing two states $s=\{1,2\}$, a uniform transition
rate $\lambda_0$ and a transition  probability matrix ${\bf A}$ given by
\begin{equation}
{\bf A}= \frac{1}{2} \left (
\begin{array}{ccc}
1+r & & 1+r \\
1-r & & 1-r
\end{array}
\right ) \, ,
\label{eq1}
\end{equation}
where $r \in [0,1)$. 
The velocities in the two states are equal in absolute value and
opposite {in sign. We denote them as 
$b(1)=b(s=1)=b_0$, $b(2)=b(s=2)=-b_0$. 
This is therefore a one-velocity model.} 
The stochastic model can be expressed as
\begin{equation}
d x(t)= b(\chi_2(t; \lambda_0, {\bf A})) \, d t \, ,
\label{eq2}
\end{equation}
where $\chi_2(t;\lambda_0,{\bf A})$ is a 2-state finite Poisson
process attaining values $\{1,2\}$, i.e., 
a Markov process characterized by the transition
rate $\lambda_0$ and  by the transition probability matrix ${\bf A}$.
The parameter $r$ in eq. (\ref{eq1}) determines a bias in the stochastic
motion. The occurrence of two stochastic states implies that
the statistical description of the process involves the
partial probability densities ${\bf p}(x,t)=(p_1(x,t),p_2(x,t))$,
and the SEO is given by $\partial {\bf p}(x,t)/\partial t= {\mathcal L}[{\bf p}(x,t)]$ where the infinitesimal generator ${\mathcal L}$ is \cite{gpk1}
\begin{equation}
{\mathcal L} =
\left (
\begin{array}{ccc}
-b_0 \frac{\partial}{\partial x} - \widetilde{\lambda} \, (1-r)
& & \widetilde{\lambda} \, (1+r) \\
 & & \\
 \widetilde{\lambda} \, (1-r) & &
b_0 \frac{\partial}{\partial x} - \widetilde{\lambda} \, (1+r)
\end{array}
\right ) \, ,
\label{eq3}
\end{equation}
and $\widetilde{\lambda}=\lambda_0/2$.
Consider the associated eigenvalue problem,
${\mathcal L}[\boldsymbol{\psi}(x)]=\mu \, \boldsymbol{\psi}(x)$.
From the structure of ${\mathcal L}$ defined by eq. (\ref{eq3}),
the eigenfunctions are
the imaginary exponentials,  $\boldsymbol{\psi}(x)= e^{i k \, x} (\psi_1^0,\psi_2^0)$.
Introducing the dimensionless quantities $\mu^*=\mu/\widetilde{\lambda}$,
$k^*=k \, b_0/\widetilde{\lambda}$, the eigenvalues of ${\mathcal L}$
can be expressed by the relation
\begin{equation}
\mu^*=-1+ \left [1- (k^*)^2 - i \, 2 \, k^* \, r \right ]^{1/2} \, ,
\label{eq4}
\end{equation}
where the square root should be interpreted in its multivalued meaning,
and $i=\sqrt{-1}$.
The eigenvalue structure {described by Eq.~\eqref{eq4}} 
is depicted in figure \ref{Fig1}.
Specifically: 
(i) the real part of the eigenvalues is lower-bounded, i.e.,
 $\mbox{Re}[\mu^*] \geq -2$ (and of course $\mbox{Re}[\mu^*]  \leq 0$). 
 For finite $\lambda_0$ this implies that
the dynamics of any perturbation {of the probability density, $\delta {\bf p}(x,t)$,} is lower-bounded as
{$||\delta {\bf p}(x,t)|| \geq C e^{-2 \,  t^*}$,}
where $|| \cdot ||$ is any suitable norm 
(for instance  the $L^2$-norm) 
and $t^*=t \widetilde{\lambda}$. 
{For this process, therefore, $\mu_l=\lambda_0$.}
 (ii) For any wavevector $k$, two spectral branches exist. 
 While in the  symmetric case,  $r=0$,
the real parts of the  spectral branches collapse into a single one
for $k^*>1$, the effect of a bias, $r>0$, is to keep the two branches separate
for any $k^*$. This phenomenon can be understood easily in the
limit $k^* \rightarrow \infty$. 
{Assuming $k^* \gg 1$ in eq. (\ref{eq4}), we can write}
$\mu^* \simeq -1 + [-(k^*)^2- i \, 2\, k^* \, r]^{1/2}$. But
$(k^*)^2+ 2 \, i \, k^* \, r= (k^*+i \, r)^2 + r^2 \simeq
(k^*+i \, r)^2$, and consequently,
$\mu^*= -1 \pm i (k^* + i \, r)$. 
This implies that the high-wavevector region of the spectrum is well
approximated by the two eigenvalue branches $\mu^*= -(1+r)+i \, k^*$ and
$\mu^*=-(1-r)-i \, k^*$, thus explaining the qualitative features
depicted in figure \ref{Fig1}.

\begin{figure}
\begin{center}
\includegraphics[width=12cm]{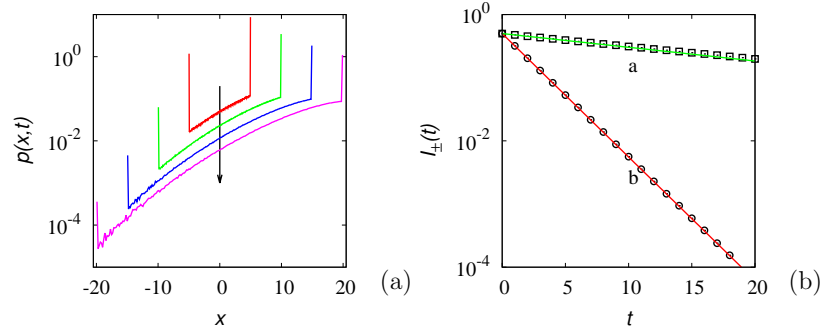}
\end{center}
\caption{(\textbf{a}) - Overall probability density $p(x,t)$ vs $x$ for
the biased 1d GPK model, with $b_0=1$, $D=1$, at $r=0.8$,
 starting from symmetric
impulsive initial conditions. The arrow indicates
increasing time instants $t=5,\,10,\,15,\,20$.
(\textbf{b}) - Decay with time of the two impulsive components $I_\pm(t)$
of the overall Green function. Symbols are the results
of stochastic simulations, lines the predictions $e^{-Re[\mu_{1,2}^\infty] \, t}$. Line (a) and ($\square$): $I_+(t)$, line (b) and ($\circ$) $I_-(t)$.}
\label{Fig2}
\end{figure}

The latter result finds a direct interpretation in terms of 
the salient properties of the Green functions for this
class of random motion. Consider the dynamics
of an ensemble of particles moving according to
eq. (\ref{eq2}) and starting from the origin.
Assume symmetric initial conditions, i.e., $p_1(x,0)=p_2(x,0)=\delta(x)/2$.
Figure \ref{Fig2} (a) depicts the evolution
of the particle ensemble, expressed in terms of the
overall probability density function, $p(x,t)=p_1(x,t)+p_2(x,t)$,
obtained from stochastic simulations of $N_p=5 \times 10^7$ particles
with $b_0=1$, $\lambda_0=1/2$, so that $D=b_0^2/2 \lambda_0=1$.
A characteristic feature 
of this class of finite-velocity stochastic
motions is that the diagonal entries of the matrix-valued Green
function are characterized by the  
 superposition of a
continuous, compactly
supported component  and of an impulsive contribution.
Expressed in terms of the overall probability  density $p(x,t)$ starting from an impulsive initial
condition, this implies that $p(x,t)=p^c(x,t)+p^\delta(x,t)$,
where $p^c(x,t)$ is the continuous term 
and  $p^\delta(x,t)$ the impulsive one, 
namely
$p^\delta(x,t)=I_+(t) \, \delta(x-b_0 \, t)+I_-(t) \delta(x+b_0 \, t)$. 
The functions $I_\pm(t)$ progressively fade away as $t$ increases \cite{kolesnik_b,giona_green,kolesnik_b1}. 
{The previous discussion suggests that}
the decay of the impulsive branches is controlled by the
real part of the two asymptotic eigenvalues $\mbox{Re}[\mu_1^\infty]=-\widetilde{\lambda}
\, (1+r)$ for $I_-(t)$; and $\mbox{Re}[\mu_2^\infty]=-\widetilde{\lambda}
\, (1-r)$ for $I_+(t)$. 
This phenomenon is depicted in figure
\ref{Fig2} (b).

{Next, we consider 
a two-dimensional GPK model possessing $N$ stochastic states.}
This process can be characterized in terms of 
$N$ partial probability densities ${\bf p}({\bf x},t)=(p_\alpha({\bf x},t))_{\alpha=1}^N$, 
satisfying the general evolution equation \cite{gpk1}
\begin{equation}
\frac{\partial p_\alpha}{\partial t}=
- {\bf b}_\alpha \cdot \nabla p_\alpha- \lambda_\alpha
\, p_\alpha + \sum_{\beta=1}^N A_{\alpha,\beta} \, \lambda_\beta \, ,
\, p_\beta
\label{eq5}
\end{equation}
with transition rates among states $\lambda_\alpha>0$, 
probability transition matrix $A_{\alpha,\beta} \geq 0$, where 
$\sum_{\alpha=1}^N A_{\alpha,\beta}=1$, for $\alpha,\,\beta=1,\dots,N$, 
and velocity vectors ${\bf b}_\alpha$.
Also in this case we are interested in  the spectral properties
of the {associate SEO (now vector-valued)} 
acting on the partial densities defined by eq. (\ref{eq5}).
The eigenfunctions are planar waves,
$\psi_\alpha({\bf x})= e^{i \, {\bf k} \cdot {\bf x}} \, \psi_\alpha^0$,
for $\alpha=1,\dots,N$.
As regards the internal structure of the GPK model, we assume
that all the transition rates are equal $\lambda_\alpha=\lambda_0$,
the transition probability matrix is  uniform $A_{\alpha,\beta}=1/N$,
and the characteristic velocity vectors are 
${\bf b}_\alpha= b_0 \, \boldsymbol{\beta}_\alpha$,
with  $\boldsymbol{\beta}_\alpha=(\cos \varphi_\alpha, \sin \varphi_\alpha)$,
and $\varphi_\alpha= 2 \, \pi(\alpha-1)/N$, for $\alpha=1,\dots,N$.
Introducing the nondimensional quantities,
${\bf k}^* = {\bf k} \, b_0/\lambda_0$, $\mu^*=\mu/\lambda_0$, 
we can write the eigenvalue equation as 
\begin{equation}
\sum_{\beta=1}^N \left [ A_{\alpha,\beta} - ( i {\bf k}^* \cdot
\boldsymbol{\beta}_\alpha +1) \, \delta_{\alpha,\beta} \right ]
\, \psi_\beta^0 = \mu^* \, \psi_\alpha^0 \, ,
\label{eq6}
\end{equation}
where $\delta_{\alpha,\beta}$ are the Kronecker symbols.

\begin{figure}
\begin{center}
\includegraphics[width=8cm]{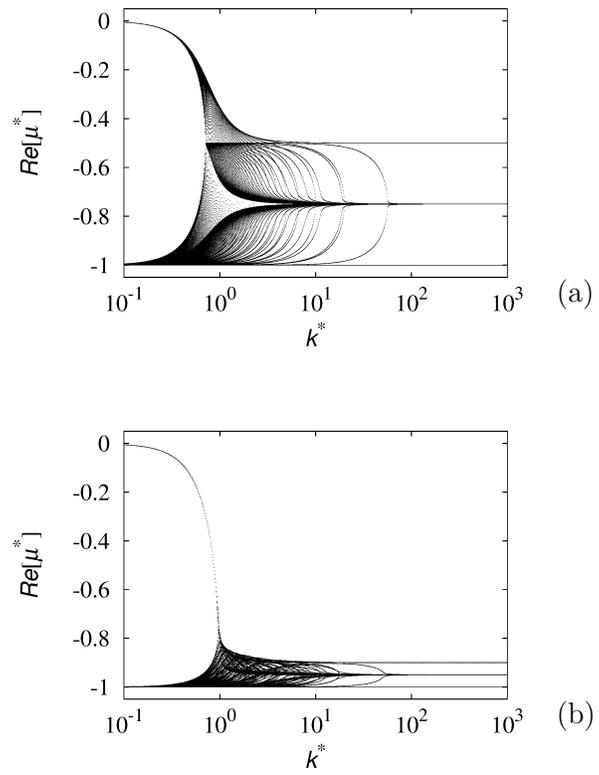}
\end{center}
\caption{Real part of the spectrum vs the wavevector $k^*$
for the 2d GPK model discussed in the text with $N$ stochastic
states. (\textbf{a}): $N=4$;  (\textbf{b}): $N=20$.}
\label{Fig3}
\end{figure}

Figure \ref{Fig3} shows the real part of the eigenvalue
spectrum of these models for two different values of $N$
as a function of the norm $k^*=|{\bf k}^*|$ of the wave vector.
The real part of the eigenvalue spectrum is lower bounded,
specifically $\mbox{Re}[\mu^*] \geq -1$. 
{For this 2d GPK process, therefore, 
we obtain the same lower bound of the 1d biased model, $\mu_l=\lambda_0$.}
The comparison of the data
shown in the two panels suggests the existence of a well defined
spectral limit for $N \rightarrow \infty$, i.e., in the
case the velocities are uniformly distributed
on the unit circumference. This limit case, corresponding to
the Markov motions analyzed by Kolesnik \cite{kolesnik}, suggests that
for a continuum of stochastic states
 the spectral properties
of the resulting SEO  give rise to a much simpler
structure of the  spectral branches.

\section{L\'evy walks}
\label{sec3}

Owing to the analysis developed by Fedotov \cite{fedotov},
subsequently elaborated in \cite{fe1,fe2,fe3,jphysa}, and extended in
a unitary theory of stochastic processes possessing bounded velocity
in \cite{epk},
 a complete  statistical
description of a LW
involves
a system of  partial probability densities, as for
GPK processes. The only difference is that
in the LW case, 
{we need to parametrize} the
partial densities  also with respect to an internal parameter, the transition
age. This is due to the more complex nature of the  density function
for the transition times that is
 no longer exponential (as for Poisson-Kac processes).
We consider a one-dimensional LW
switching between two states, corresponding to the directions
of motion, keeping constant the absolute value of the velocity $b_0$.
The statistical properties of such a LW
are fully described by the partial probability densities $p_\pm(x,t;\tau)$.
Setting
\begin{equation}
{\bf p}(x,t;\tau)=
\left (
\begin{array}{c}
p_+(x,t;\tau) \\
p_-(x,t;\tau)
\end{array}
\right ) \, ,
\label{eq6_1}
\end{equation}
the vector-valued  density ${\bf p}(x,t;\tau)$
satisfies the evolution equation \cite{fedotov}
{\begin{equation}
\frac{\partial {\bf p}}{\partial t}= {\mathcal M}_\tau [{\bf p}]+ {\mathcal A}_x [
{\bf p} ]
={\mathcal L}_{\tau,x}[ {\bf p}] \, ,
\label{eq6_2}
\end{equation}
where ${\mathcal M}_\tau$ represents the evolution operator for the transition age,  
\begin{equation}
{\mathcal M}_\tau = 
\left (
\begin{array}{cc}
-\frac{\partial }{\partial \tau}
-\lambda(\tau) & 0 \\
0 & - \frac{\partial }{\partial \tau}
- \lambda(\tau)
\end{array}
\right ) = - {\bf I}_2 \left(\frac{\partial }{\partial \tau}
+\lambda(\tau) \right) \, ,
\label{eq6_2a}
\end{equation}
${\bf I}_2 $ is the $2\times2$ identity matrix, and}
${\mathcal A}_x$ represents the
advection operator
\begin{equation}
{\mathcal A}_x =
\left (
\begin{array}{cc}
- b_0 \, \frac{\partial }{\partial x} & 0 \\
0 & b_0 \, \frac{\partial }{\partial x}
\end{array}
\right ) = - \boldsymbol{\sigma}_z \, b_0 \,  \frac{\partial}{\partial x} \, ,
\label{eq6_3}
\end{equation}
with $\boldsymbol{\sigma}_z = \left ( \begin{array}{cc} 1 & 0 \\ 0 & -1 \end{array}
\right )$ a Pauli matrix.
{Both operators act on the two partial probability density waves $p_{\pm}$.}
Assuming that at each transition,  particles invert their direction of motion,
eq. (\ref{eq6_2}) is equipped with the boundary condition
\begin{equation}
{\bf p}(x,t,0)=  \boldsymbol{\sigma}_x \, \int_0^\infty \lambda(\tau) \,
{\bf p}(x,t,\tau) \, d \tau \, ,
\label{eq6_4}
\end{equation}
where $\boldsymbol{\sigma}_x$ is the Pauli matrix
$\boldsymbol{\sigma}_x= \left ( \begin{array}{cc} 0 & 1 \\ 1 & 0 \end{array}
\right )$.
We now consider the spectral structure underlying the SEO, namely
we solve the eigenvalue problem,
\begin{equation}
{\mathcal L}_{\tau,x}[\boldsymbol{\psi}(x,\tau)] = \mu \, {\boldsymbol \psi}(x,\tau) \; , \qquad
{\boldsymbol \psi}(x,\tau) =
\left (
\begin{array}{c}
\psi_+(x,\tau) \\
\psi_-(x,\tau)
\end{array}
\right ) \, .
\label{eq6_5}
\end{equation}
The eigenfunction ${\boldsymbol \psi}(x,\tau)$ can be
expressed as
planar waves with respect to the spatial coordinate, i.e., 
${\boldsymbol \psi}(x,\tau) = e^{i \, k \, x} \boldsymbol{\phi}(\tau)$,
while the dependence on the transition time $\tau$ is accounted for
by the vector-valued function $\boldsymbol{\phi}(\tau)=(\phi_+(\tau),\phi_-(\tau))$.
Its two components satisfy the equations
\begin{equation}
\frac{d \phi_\pm(\tau)}{d \tau}  =
- \left ( \mu \pm i \, k \, b_0 + \lambda(\tau) \right ) \,
\phi_\pm(\tau) \, ,
\label{eq6_6bis}
\end{equation}
and thus,
\begin{equation}
\phi_\pm(\tau)  =  A_\pm \, \exp \left [-\mu \, \tau \mp i \, k \, b_0\ \, \tau
- \int_0^\tau \lambda(\theta) \, d \theta \right ] \, ,
\label{eq6_7}
\end{equation}
where $A_\pm$ are two complex-valued constants.

{Imposing the boundary conditions (\ref{eq6_4}) on the function $\boldsymbol{\phi}$, 
we find that the two constants $A_\pm$ satisfy the relation}
\begin{equation}
A_\pm  =   A_{\mp} \,
\int_0^\infty T(\tau) e^{-(\mu \pm i \, k \, b_0) \, \tau} \, d \tau \, ,
\label{eq6_8}
\end{equation}
where $T(\tau)=\lambda(\tau) \, \exp \left
 [-\int_0^\tau \lambda(\theta) \, d \theta
\right ]$
is the probability density function for the transition times.
{Eliminating $A_{\pm}$ from the two eqs. (\ref{eq6_8}), we obtain} 
that the characteristic
equation for the eigenvalues of ${\mathcal L}_{\tau,x}$ is
given by
\begin{equation}
\left [
\int_0^\infty T(\tau) e^{-(\mu+i \, k \, b_0) \, \tau} \, d \tau
\right ]   \left [\int_0^\infty T(\tau^\prime) e^{-(\mu-i \, k \, b_0) \, \tau^\prime} \, d \tau^\prime
\right ] = 1 \, .
\label{eq6_9}
\end{equation}
For $k=0$ one recovers the spectral properties of
the ``renewal mechanism'' of the LW. In this
case, eq. (\ref{eq6_9}) reduces to
\begin{equation}
\left [ \int_0^\infty T(\tau) e^{-\mu \, \tau} \, d \tau \right ]^2
= 1 \, .
\label{eq6_10}
\end{equation}
Equation (\ref{eq6_10}) can be interpreted as follows:
the spectral structure of the renewal equation of
a LW corresponds to the zeroes of the equations
$\widehat{T}(\mu) = \pm 1$
where  $\widehat{T}(\mu)$  represents the Laplace transform
(of argument $\mu$) of the transition-time probability
density $T(\tau)$.
For real $\mu$, only the positive determination of the latter equation
makes sense, and it is straightforward to notice that there
exists a unique real solution of $\widehat{T}(\mu) = 1$,
namely $\mu=0$,
corresponding to the Frobenius  (conservation) eigenvalue
of the process.

{In the rest of this section, we consider three typical LW models and solve for their spectral properties.}

\subsection{Gamma-distributed transition times}
\label{sec3.1}

For this model, the transition-time density is expressed by
\begin{equation}
T(\tau) = \frac{\beta^\alpha}{\Gamma(\alpha)} \, \tau^{\alpha-1} \, e^{-\beta \, \tau} \,,
\label{eq6_17}
\end{equation}
with $\alpha \in (0,\infty)$, $\beta>0$. Since
\begin{equation}
\widehat{T}(s) = \frac{\beta^\alpha}{(s+\beta)^\alpha} \, ,
\label{eq6_18}
\end{equation}
it follows  that eq. (\ref{eq6_9}) reduces to
\begin{equation}
\left [ (\mu+\beta)^2 + k^2 \, b_0^2 \right ]^\alpha=\beta^{2 \alpha} \, ,
\label{eq6_19}
\end{equation}
{or equivalently,} 
\begin{equation}
(\mu+\beta)^2 + k^2 \, b_0^2 = \beta^2 e^{i \, \varphi_{\alpha,h}} \, ,
\label{eq6_23}
\end{equation}
where
$\varphi_{\alpha,h}=2 \, \pi \, h/\alpha$, and $h$ is  an integer,
 $h=0,1,\dots$. Two cases may occur: (i) if
 $\alpha=P/Q$ is rational, with $P,\,Q$ both integers,
there are $Q$ distinct values of $e^{i \, \varphi_{\alpha,h}}$ corresponding
to $h=0,1,\dots,Q$; (ii) if $\alpha$ is irrational, 
{there is a countable number of distinct values of $e^{i \, \varphi_{\alpha,h}}$, 
each corresponding to a different spectral branch.}

The spectral properties
of the renewal mechanism (corresponding to $k=0$)
follow from eq. (\ref{eq6_23})
\begin{equation}
\mu_h= - \beta \pm \beta e^{i \, \varphi_{\alpha,h}/2} \, ,
\label{eq6_24}
\end{equation}
and  the eigenvalues are in general
complex-valued. In particular, if $\alpha$ is irrational,
 $(\mu_h+\beta)/\beta$,
$h=0,1,\dots$, fill densely the unit circumference.

Next, we consider the case $k\neq 0$. Solving the quadratic equation
(\ref{eq6_23}) one obtains
\begin{equation}
\mu_h= - \beta \pm \left [\beta^2 e^{i \, \varphi_{\alpha,h}}-k^2 \, b_0^2
\right ]^{1/2} = - \beta \pm \sqrt{z_h} \, ,
\label{eq6_25}
\end{equation}
where $z_h= \left [ \beta^2 \cos(\varphi_{\alpha,h}) - k^2 \, b_0^2 \right ]
+ i \, \beta^2 \, \sin(\varphi_{\alpha,h})= \rho_h \, e^{i \, \omega_h}$.
Since the two determinations of $\sqrt{z_h}$ are $\pm \rho_h^{1/2} e^{i \,
\omega_h/2}$, it follows that
\begin{equation}
\mu_h = - \beta \pm \rho_h^{1/2} \, \left [
\cos \left (\frac{\omega_h}{2} \right ) + i \, \sin \left (\frac{\omega_h}{2} \right ) \right ] \, .
\label{eq6_26}
\end{equation}
It is straightforward to show that the real part of the
spectrum is lower bounded. Moreover, the explicit calculation
of the eigenvalues
indicates that
\begin{equation}
- 2 \, \beta \leq \mbox{Re}[\mu_h] \leq 0 \, .
\label{eq6_26bis}
\end{equation}

\begin{figure}
\begin{center}
\includegraphics[width=8cm]{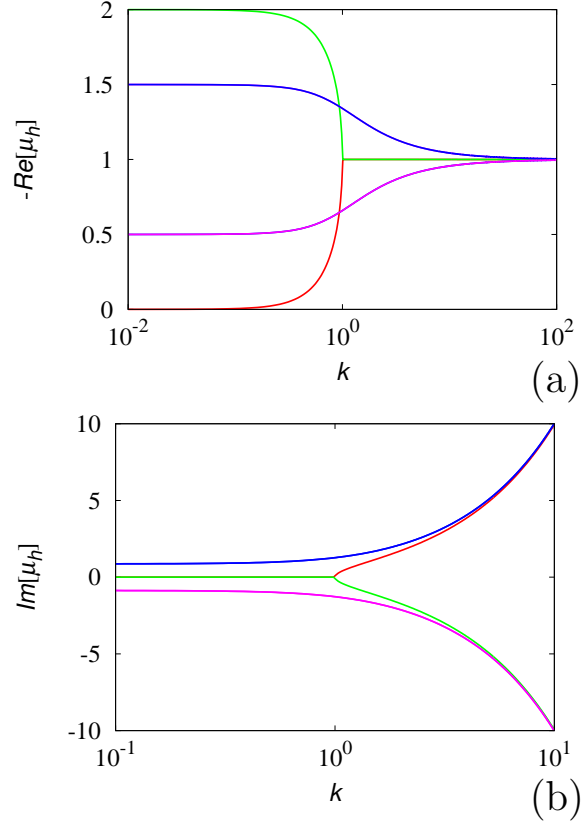}
\end{center}
\caption{Spectral structure of the Gamma-distributed LW at $\alpha=3$,
$\beta=b_0=1$ a.u.  (\textbf{a}): $-\mbox{Re}[\mu_h]$ vs $k$; (\textbf{b}):
$\mbox{Im}[\mu_h]$ vs $k$.}
\label{Fig4}
\end{figure}

\begin{figure}
\begin{center}
\includegraphics[width=8cm]{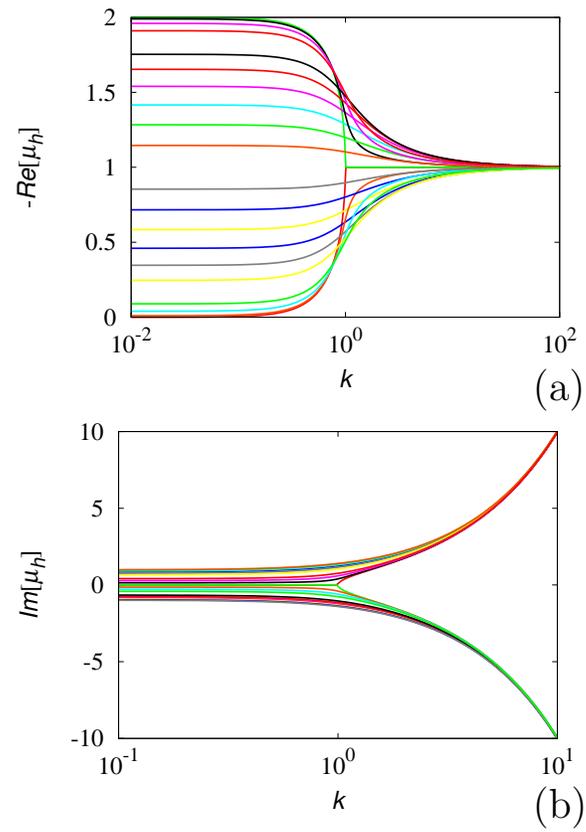}
\end{center}
\caption{Spectral structure of the Gamma-distributed LW at $\alpha=\pi$,
$\beta=b_0=1$ a.u.  (\textbf{a}): $-\mbox{Re}[\mu_h]$ vs $k$;  (\textbf{b}):
$\mbox{Im}[\mu_h]$ vs $k$. Solely the first 20 spectral branches are depicted.
}
\label{Fig5}
\end{figure}

In order to give some examples, figures \ref{Fig4} and \ref{Fig5}
depict the real and imaginary part of the spectrum in two
typical cases: $\alpha=3$ (integer), and $\alpha=\pi$ (irrational).
The other parameters are set to $\beta=b_0=1$.
In the irrational case, solely the the spectral branches
corresponding to $h=0,\dots,19$, are depicted.

\subsection{Superdiffusive L\'evy walk}
\label{sec3.2}

{For  LWs   defined
by a transition-time probability density possessing a long-term power-law
scaling, such as  those associated with the distribution
function $ F_\tau(\tau) = \int_0^\tau T(\tau^\prime) \, d \tau^\prime = 1-E_\nu(-(\tau/\tau_0)^\nu)$,
 where $E_\nu(z)= \sum_{k=0}^\infty z^k/\Gamma(\nu \, k+1)$ is
  the Mittag-Leffler function, $0<\nu<1$,  $\tau_0>0$,  and $\Gamma(z)$ the Gamma function \cite{scalas, sheng},
the Laplace transform of
$T(\tau)$ is
given by}
\begin{equation}
\widehat{T}(s)= \frac{1}{1+(\tau_0 \, s)^\nu} \, ,
\label{eq6_20}
\end{equation}
and eq. (\ref{eq6_9}) provides
\begin{equation}
\left [1+ \tau_0^\nu \, (\mu+i \, k \, b_0)^\nu \right ] \,
\left [1+ \tau_0^\nu \, (\mu-i \, k \, b_0)^\nu \right ] =1 \, .
\label{eq6_21}
\end{equation}



To solve this equation, we set
\begin{equation}
\mu + i \, k \, b_0 = \mbox{Re}[\mu]+ i \left ( \mbox{Im}[\mu] \pm k \, b_0 \right )
= \, \rho \, e^{i \phi_\pm} \, ,
\label{eq9_10}
\end{equation}
where $\rho  \geq \mbox{Re}[\mu]$.
Eq. (\ref{eq6_21}) can be thus rewritten as
\begin{equation} 
1 +  \tau_0^\nu \, \rho^\nu \, \left ( e^{i \, \nu \, \phi_+} +
e^{i \, \nu \, \phi_-} \right ) +
\tau_0^{2 \, \nu} \rho^{2 \, \nu} \, e^{i \, \nu \, (\phi_+ + \phi_-)}
=1 \, , 
\end{equation}
which can be further reduced to
\begin{equation}
\tau_0^\nu \, \rho^\nu + (e^{-i \, \nu \, \phi_+}+e^{-i \nu \phi_-})
=0 \, . 
\label{eq9_12}
\end{equation}
This implies the relation 
\begin{equation}
|\mbox{Re}[\mu]| \leq \rho \leq \frac{2^{1/\nu}}{\tau_0} \, ,
\label{eq9_13}
\end{equation}
which proves the boundedness of the real part of the spectrum. 

\subsection{Power-law distributed transition times}
\label{sec3.3}

We consider as a third example the classical case of a LW (\ref{eq6_2})-(\ref{eq6_4}) defined by the
the transition-time   probability density $T(\tau)=\xi/(1+\tau)^{\xi+1}$, where
$\xi>0$.
{Its Laplace transform is given by
\begin{equation}
\widehat{T}(\mu) = \xi \, \int_0^\infty \frac{e^{-\mu \, \tau}}{(1+\tau)^{\xi+1}}
\, d \tau=  \xi \, e^{\mu} \, \mbox{Ei}_{\xi+1}(\mu) \, ,
\label{eq9_1}
\end{equation}
where $\mbox{Ei}_{\xi+1}(z)$ is the Exponential Integral function of
order $\xi+1$, 
\begin{equation}
\mbox{Ei}_\nu(z)= \int_1^\infty \frac{e^{-z y}}{y^\nu} \, d y \, ,
\label{eq9_2}
\end{equation}
}{In this case, it is not possible to derive in closed-form the structure of the eigenvalue branches.
Nevertheless, we can still obtain spectral bounds using asymptotic analysis. }

Let us prove {\em ad absurdum} that there exists a lower
bound for the real part of the eigenvalue
spectrum. Let us suppose the opposite, namely that
there are eigenvalues with arbitrarily large real part,
that implies arbitrarily large values of $|\mu|$.
In this case, we can use eq. (\ref{eq9_1}), and the
asymptotic expansion for the Exponential Integral
{function $\mbox{Ei}_\nu(z)$ for complex $z$ \cite{table},
\begin{equation}
\mbox{Ei}_\nu(z)= \frac{e^{-z}}{z} \left [ 1 - \frac{\nu}{z} + O \left ( \frac{1}{z^2}
\right ) \right ] \, ,
\label{eq9_3}
\end{equation}
}
which implies
\begin{equation}
\widehat{T}(\mu \pm i \, k \, b_0) = \frac{\xi}{ \mu \pm i \, k \, b_0}
+ \left .  O \left ( \frac{1}{s^3}  \right ) \right |_{s=\mu \pm i \, k \, b_0} \, .
\label{eq9_4}
\end{equation}
For large $\mbox{Re}[\mu]$, one obtains from eq. (\ref{eq6_9})
\begin{equation}
\frac{\xi^2}{(\mu+ i \, k \, b_0) \, (\mu-i \, k \, b_0)} =1 +
O \left ( \frac{1}{\mbox{Re}[\mu]^4} \right ) \, .
\label{eq9_5}
\end{equation}
By hypothesis, the real part of the eigenvalues can attain arbitrarily
large values so that the higher-order terms in eq. (\ref{eq9_5}) can be
neglected, reducing it to the equation
\begin{equation}
\mu^2 + k^2 \, b_0^2 - \xi^2 =0  \, . 
\label{eq9_6}
\end{equation}
The solutions of this equation are
\begin{equation}
\mu=
\left \{
\begin{array}{lll}
\pm \sqrt{\xi^2- k^2 \, b_0^2} & & |k| < \xi/b_0 \\
\pm i \, \sqrt{ k^2 \, b_0^2-\xi^2} & & |k|>\xi/b_0 \, .
\end{array}
\right .
\label{eq9_7}
\end{equation}
It  follows from eq. (\ref{eq9_7})  that the real part
of the spectrum is lower-bounded by $-\xi/b_0$, contradicting
the original hypothesis. The proof is completed.

From the properties of Laplace transforms, namely $\lim_{{\mbox Re}[\mu]
\rightarrow \infty} \widehat{T}(\mu)=0$, and applying the same analysis
developed above, we can show that all the one-dimensional
LWs, whose spectrum fulfils eq. (\ref{eq6_9}), are
characterized by a lower bound $-\mu_l$, with $\mu_l>0$ for
$\mbox{Re}[\mu]$.

\section{GPK process with a continuum of states}
\label{sec4}

A common feature of one-velocity models is the decomposition
of the Green function into a continuous and an impulsive part
as discussed in \cite{kolesnik_b,giona_green} and shown pictorially
in figure \ref{Fig2}.
 
A question naturally arises, whether it would be possible
to define a stochastic process with finite propagation velocity, for which
the corresponding  matrix-valued Green function does not 
admit an impulsive contribution but solely the continuous part,
similarly to what happen for the solution of the parabolic diffusion
equation. While the answer to this question is negative (for reasons
that are addressed below), it is indeed possible to
construct processes such that, if the overall initial
density is impulsive, say
$p(x,0)=\delta(x)$,
then for any $t>0$ the
overall density function
$p(x,t)$ is  smooth and compactly supported,
i.e., it does not not possess any impulsive term. This
construction is interesting for further addressing the spectral properties
of these models,  as the sudden annihilation of the impulsive initial
 conditions necessarily  implies a spectral
counterpart in the high-wavevector limit.
{Examples of these models have been discussed in \cite{zaburdaev2016}.}

Processes possessing this property implies
a continuous set of stochastic states possessing
velocities defined in some bounded and continuous set (in the present case an interval),
and smooth transitions amongst them. In this
case, the velocity itself can be used to parametrize
the stochastic states.
A typical example is given by
the stochastic dynamics
\begin{equation}
d x(t) = b_0 \, \Xi(t;\lambda; K) \, d t \, ,
\label{eqx_1}
\end{equation}
where $\Xi(t;\lambda,K)$ is a Poisson field defined in \cite{gpk3},
and attaining values in $[-1,1]$, i.e., a stochastic
process such that the probabilities densities $\widehat{P}(\beta,t)$
where $\widehat{P}(\beta,t) \, d \beta=
\mbox{Prob}[\Xi(t) \in (\beta,\beta+d \beta)]$, satisfy
a Markov dynamics
\begin{equation}
\frac{\partial \widehat{P}(\beta,t)}{\partial t} = -\lambda(\beta) \, \widehat{P}(\beta,t)
+ \int_{-1}^1 K(\beta,\beta^\prime) \, \lambda(\beta^\prime) \, \widehat{P}(\beta^\prime,t) \,  d \beta^\prime \, ,
\label{eqx_2}
\end{equation}
where $\lambda(\beta)$ is a smooth positive function expressing the
transition rate from the stochastic state $\beta$, and 
$K(\beta,\beta^\prime)$ a smooth stochastic kernel,
with $K(\beta,\beta^\prime)  \geq 0$, and
\begin{equation}
\int_{-1}^1 K(\beta,\beta^\prime) d \beta = 1 \, ,
\label{eqx_3}
\end{equation}
for all $\beta^\prime \in [-1,1]$. 
The latter accounts for the transition
probabilities from state $\beta^\prime$ to all the other states.
Henceforth, we consider the simplest case of uniform $\lambda(\beta)$
and $K(\beta,\beta^\prime)$, i.e., 
\begin{equation}
\lambda(\beta)=\lambda_0 \, , 
\qquad K(\beta,\beta^\prime) = \frac{1}{2} \, .
\label{eqx_4}
\end{equation}
The statistical description of the stochastic  motion
defined by eq. (\ref{eqx_1}) involves the family of partial
probability densities $p(x,t;\beta)$, continuously
parametrized with respect to $\beta \in [-1,1]$,
and satisfying the equation
\begin{equation}
\frac{\partial p (x,t;\beta)}{\partial t} = - b_0 \, \beta \, \frac{\partial p(x,t;\beta)}{\partial x}
- \lambda_0 \, p(x,t;\beta) + \frac{\lambda_0}{2} \int_{-1}^1
p(x,t;\beta^\prime) \, d \beta^\prime \, .
\label{eqx_5}
\end{equation}
In this model, the overall probability density $P(x,t)$ for the particle
position $x$ at time $t$, and the associated flux $J(x,t)$ are
given by
\begin{equation}
P(x,t)= \int_{-1}^1 p(x,t;\beta) \, d \beta \, , \qquad
J(x,t)= \int_{-1}^1 \beta \,  p(x,t;\beta) \, d \beta \, .
\label{eqx_6}
\end{equation}
Using the method outlined in \cite{gpk1,gpk2,gpk3}, it follows that
the Kac limit of eq. (\ref{eqx_5}), i.e., the limit for
unbounded propagation velocity and transition rate,
 $b_0,\, \lambda_0 \rightarrow \infty$,
keeping fixed the nominal diffusivity $D_{\rm nom}=b_0^2/2 \, \lambda_0$
is given by the diffusion equation for $P(x,t)$,
\begin{equation}
\frac{\partial P(x,t)}{\partial t}= D \, \frac{\partial^2 P(x,t)}{\partial x^2} \, ,
\label{eqx_7}
\end{equation}
with a value of the effective diffusivity $D$ equal to
$D= 2 \, D_{\rm nom}/3$. Moreover, for any finite value
of $b_0$ and $\lambda_0$, the long-term solutions of eq. (\ref{eqx_5})
approach those of the parabolic equation (\ref{eqx_7}).

\begin{figure}
\begin{center}
\includegraphics[width=8cm]{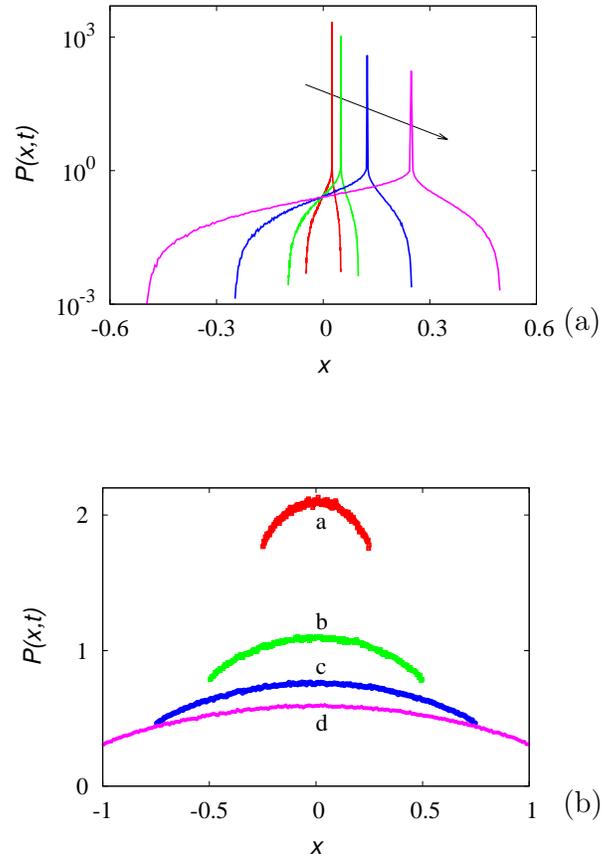}
\end{center}
\caption{Overall probability density function $P(x,t)$ vs $x$
associated with the process  defined by eqs. (\ref{eqx_1}) and (\ref{eqx_4})
at $b_0=1$, $\lambda_0=1/2$, with $P(x,0)=\delta(x)$,
 for two different initial
preparations.  (\textbf{a}) refers to an impulsive initial
distribution $\widetilde{p}_0(\beta)=\delta(\beta-0.5)$,
and the arrow indicates increasing values of time $t=0.05,\, 0.1,\,0.25,\, 0.5$.
 (\textbf{b}) refers to a smooth and uniform initial preparation
$\widetilde{p}_0(\beta)=1/2$. Points (a) correspond to $t=0.25$, (b) to $t=0.5$,
(c) to $t=0.75$, (d) to $t=1.0$.}
\label{Fig6}
\end{figure}

Consider the solutions of eq. (\ref{eqx_5}) starting from
a spatially impulsive initial distribution $P(x,0)=\delta(x)$.
This condition does not specify completely the initial state
of the system, as the  initial preparation with respect to the
internal parametrization $\beta$ should be also defined, i.e.,
the whole structure of the 
 partial densities $p(x,0;\beta)=\widetilde{p}_0(\beta) \, \delta(x)$, $\beta \in [-1:1]$ at $t=0$ should be specified.
Two situation can occur: (i) if $\widetilde{p}_0(\beta)=\delta(\beta-\beta^*)$ 
admits an impulsive
component  at any $\beta=\beta^*$, then $P(x,t)$ for $t>0$ is characterized
by an impulsive contribution centered at $x=\beta^* \, t$,   propagating
with velocity $\beta^*$
superimposed to a continuous distribution deriving from the recombination mechanism amongst the partial waves.
Conversely, (ii) if $\widetilde{p}_0(\beta)$ is smooth, i.e., no
impulsive initial term is present, then
$P(x,t)$ for $t>0$ is also a smooth function of $x$ for  any $t>0$.
The phenomena outlined above are depicted in figure
\ref{Fig6} panels (a) and (b). Data have been obtained
from stochastic simulations unsing an ensemble of $N_p=5 \times 10^7$ walkers
initially placed at $x=0$. In the case of the
data depicted in panel (a)  the initial distribution with respect to
the internal state variable is
impulsive and centered at $\beta^*=1/2$. Conversely, the
initial condition for the data depicted in panel (b) is
uniform over $\beta$, i.e., $\widetilde{p}_0(\beta)=1/2$, $\beta \in [-1,1]$.

This example clearly indicates that even in the presence of a continuum
of internal states, any initial preparation that is impulsive with
respect to the internal-state parametrization $\beta$ determines
an overall density $P(x,t)$ containing an impulsive contribution
with respect to $x$, propagating at constant speed. 
It is interesting to analyze how this phenomenology can be interpreted
in the light of the spectral properties of the associated
evolution operator.

\subsection{Spectral properties}
\label{sec5}
\begin{figure}
\begin{center}
\includegraphics[width=8cm]{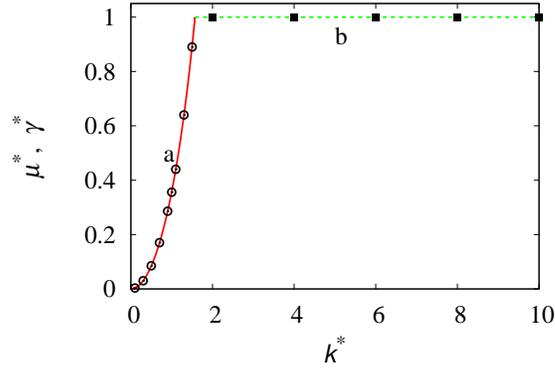}
\end{center}
\caption{Eigenvalue branch $\mu^*$ vs the wavenumber $k^*$ (line a).
Line (b) represents the maximum relaxation exponent $\gamma^*=1$, associated
with the essential spectrum of the evolution operator. Symbol ($\circ$) and ($\blacksquare$) are the result of the numerical simulations of eq. (\ref{eq3_11}),
considering the decay of the $L^2$-norm of the solution.}
\label{Fig7}
\end{figure}

\begin{figure}
\begin{center}
\includegraphics[width=8cm]{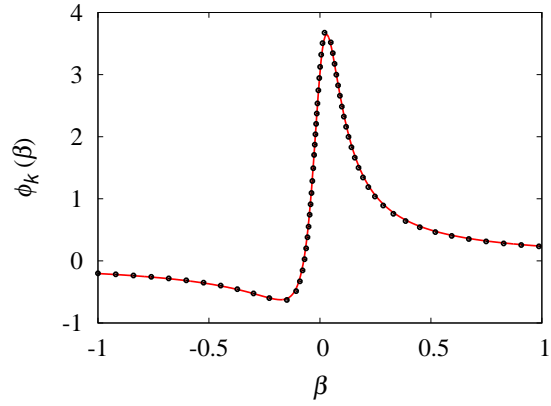}
\end{center}
\caption{Profile of the  real part of the eigenfunction $\phi_k(\beta)$ vs $\beta$ at $k^*=1.5$. The line represents the graph of eq. (\ref{eq3_10}), symbols ($\bullet$) the profile
obtained from the relaxation of eq. (\ref{eq3_11}), integrated numerically.}
\label{Fig8}
\end{figure}

Consider the spectral properties of the
infinitesimal generator of the
density dynamics (\ref{eqx_5}),
\begin{equation}
{\mathcal L}_{\beta}[\phi(x,\beta)]=
- b_0 \, \beta \, \frac{\partial \phi(x,\beta)}{\partial x}
- \lambda_0 \, \phi(x,\beta) + \frac{\lambda_0}{2} \int_{-1}^1
\phi(x,\beta^\prime) \, d \beta^\prime \, .
\label{eq3_1}
\end{equation}
The eigenvalue equation for ${\mathcal L}_{\beta}$, 
\begin{equation}
{\mathcal L}_\beta[\psi(x,\beta)]=\mu \, \psi(x,\beta)  \, ,
\label{eq3_2}
\end{equation}
admits the eigenfunctions of the form
\begin{equation}
\psi(x,\beta)= e^{i \, k \, x} \, \psi_k(\beta) \, .
\label{eq3_3}
\end{equation}
Introducing the dimensionless quantities $\mu^*=\mu/\lambda_0$,
$k^*=k \, b_0/\lambda_0$, the 
substitution of eq. (\ref{eq3_3})
 into eqs. (\ref{eq3_1})-(\ref{eq3_2}) provides
\begin{equation}
\psi_k(\beta) = \frac{1}{2 \, (1+\mu^* + i \, k^* \, \beta)}
\int_{-1}^1 \psi_k(\beta^\prime) \, d \beta^\prime \, .
\label{eq3_4}
\end{equation}
Consequently, integrating over $\beta$ and assuming 
$\int_{-1}^1 \psi_k(\beta) \, d \beta \neq 0$, 
the characteristic equation for the eigenvalues  follows
\begin{equation}
\int_{-1}^{1} \frac{d \beta}{1+\mu^* + i k^* \, \beta} = 2 \, .
\label{eq3_5}
\end{equation}
Setting $\mu^*=r + i \, \omega$, 
and expliciting
the integral in eq. (\ref{eq3_5}) one obtains 
the equation
\begin{equation}
\mbox{arctan} \left  ( \frac{\omega +k^*}{1+r} \right )
- \mbox{arctan} \left  ( \frac{\omega -k^*}{1+r} \right )
- \frac{i}{2} \mbox{log} \left [ \frac{(1+r)^2 + (\omega+ k^*)^2}
{(1+r)^2 + (\omega - k^*)^2} \right ] = 2 \, k^* \, .
\label{eq3_6}
\end{equation}
{Imposing the imaginary part to be null}
one obtains $\omega=0$, i.e., the eigenvalues are purely real, 
$\mu^*=r$. Enforcing this property back in eq.  
(\ref{eq3_6}), a simpler equation for the real part $r$ follows, 
\begin{equation}
k^* = \mbox{arctan} \left ( \frac{k^*}{1+r} \right ) \, .
\label{eq3_7}
\end{equation}
The solution of eq. (\ref{eq3_7}) can be easily obtained
by introducing the auxiliary variable $z=k^*/(1+r)$, so that eq. (\ref{eq3_7}) becomes
$(1+r) z = \mbox{arctan}(z)$. Varying   $z \in [0,\infty)$, one
gets the continuous branch of eigenvalues
\begin{equation}
r= - 1+ \frac{\mbox{arctan}(z)}{z} \, .
\label{eq3_8}
\end{equation}
Given $z$, the corresponding wavenumber $k^*$ follows from
the definition of $z$, namely $k^*=(1+r) \, z$.
Before addressing the properties of this
spectral branch, let us show that there is no other
eigenvalue. To show  this we have to consider the
only remaining condition, namely that 
\begin{equation}
\int_{-1}^1 \psi_k(\beta) \, d \beta=0 \, .
\label{eq3_9}
\end{equation}
In this case, the eigenvalue problem reduces to
$(\mu^* +1 +i \, k^* \, \beta ) \, \psi_k(\beta)=0$, which
admits no solution for a $\mu^*$ independent of $\beta$.
Consequently, the spectrum reduces to the
spectral branch defined  by eqs. (\ref{eq3_7})-(\ref{eq3_8}).
This branch is defined for $|k^*|<k_c=\pi/2$, as depicted
in figure \ref{Fig7}. 
In this interval of wavevectors,
 the associated eigenfunctions  defined by
 eq. (\ref{eq3_4}) are complex-valued and
can be expressed as
\begin{equation}
\psi_{k}(\beta)= A \left [\phi_k(\beta)+ i \phi_k(-\beta) \right ] \, ,
\label{eq3_10a}
\end{equation}
where $A$, is a  complex-valued constant and $\phi_k(\beta)$
is a real-valued function  defined as 
\begin{equation}
\phi_k(\beta)= C \, \frac{(1+\mu^*) + k^* \, \beta}{(1+\mu^*)^2 + (k^* \, \beta)^2} \, ,
\label{eq3_10}
\end{equation}
where the normalization constant $C$ is such that
$|| \phi_k(\beta)||_{L^2}= \int_{-1}^1 \phi_k^2(\beta) \, d \beta=1$.
Expression (\ref{eq3_10a}) with $\phi_k(\beta)$ given by
eq. (\ref{eq3_10})  follows from eq. (\ref{eq3_4}), i.e.,
$\psi_k(\beta)= B/[(1+\mu^*)+i \, k^* \, \beta]$, setting
$B=1+i$, modulo an arbitrary multiplicative constant.
Figure \ref{Fig8} depicts the shape of $\phi_k(\beta)$ at
$k^*=1.5$, close to the break-up point $k_c \simeq 1.57$.
From the functional form of the eigenfunctions (\ref{eq3_10a})-(\ref{eq3_10}),
the dynamical mechanism associated with the spectral break-up becomes clear.
As the eigenvalue $\mu^*$ approaches $-1$ from below, the
associated eigenfunction develops a singularity at $\beta=0$,
$\psi_k(\beta) \rightarrow 1/\beta$, and consequently the
critical point $k_c$ corresponds to the bifurcation point
where $\psi_k(\beta)$ ceases to be summable.

The occurrence of a single eigenfunction, restricted solely to the
interval $[-k_c,k_c]$ is manifestly incomplete in order to represent a
generic (square-summable) function of the internal variable $\beta$ in
the interval $[-1,1]$.  It is therefore interesting to analyze the
relaxation properties of the operator ${\mathcal L}_\beta$, restricted
to a planar wave mode $p(x,t;\beta)=e^{i \, k \, x} \,
\zeta(\beta,t)$, i.e., to consider the dynamics
{
\begin{equation}
\frac{\partial \zeta(\beta,t)}{\partial  t}= -i \, k^* \, \beta \, \zeta(\beta,t)
-\zeta(\beta,t) + \frac{1}{2} \int_{-1}^1 \zeta(\beta^\prime,t) \, d \beta^\prime  \, ,
\label{eq3_11}
\end{equation}
where $\zeta(\beta,t)$ is complex valued, and the time variable $t$
has been made nondimensional   normalizing it by $\lambda_0$.
It follows from eq. (\ref{eq3_11}) that
\begin{eqnarray}
\frac{\partial |\zeta(\beta,t)|^2}{\partial t} &  = & \overline{\zeta}(\beta,t) \, \frac{\partial
\zeta(\beta,t)}{\partial t} + \zeta(\beta,t) \, \frac{\partial \overline{\zeta}(\beta,t) }{\partial t}
 = - 2 \, |\zeta(\beta,t)|^2 
\nonumber \\
&+ & \mbox{Re} \left [ \overline{\zeta}(\beta,t) \, \int_{-1}^1
\zeta(\beta^\prime,t) \, d \beta^\prime \right ] \, ,
\label{eq3_12}
\end{eqnarray}
}
where $\mbox{Re}[\cdot ]$ is the real part of the argument, and
$\overline{\zeta}(\beta,t)$ the complex conjugate of $\zeta(\beta,t)$.
Integrating over $\beta$, the dynamics of the
$L^2$-norm of $\zeta(\beta,t)$, $||\zeta||^2(t)= \int_{-1}^1 |\zeta(\beta,t)|^2 \, d \beta$, follows
\begin{equation}
\frac{d ||\zeta||^2(t)}{d t}= - 2 \, ||\zeta||^2(t) +|Z(t)|^2 \, ,
\label{eq3_13}
\end{equation}
where $Z(t)= \int_{-1}^1 \zeta(\beta,t) \, d \beta$. As  $|Z(t)|^2 \geq 0$,
it follows from eq. (\ref{eq3_13}) that
\begin{equation}
\frac{d ||\zeta||^2(t)}{d t} \geq -2 \, ||\zeta||^2(t) 
\qquad \Rightarrow \qquad
 ||\zeta||(t) \geq  ||\zeta||(0) \, e^{-t} \, .
\label{eq3_14}
\end{equation}
Eq. (\ref{eq3_14}) indicates that all the planar-wave modes
cannot decay to zero faster than $e^{-\gamma^* \, t}$,
with $\gamma^*=1$ representing the maximum relaxation rate.
This result is valid for any $k^*$, below and above the critical
value $k_c$.

\begin{figure}
\begin{center}
\includegraphics[width=12cm]{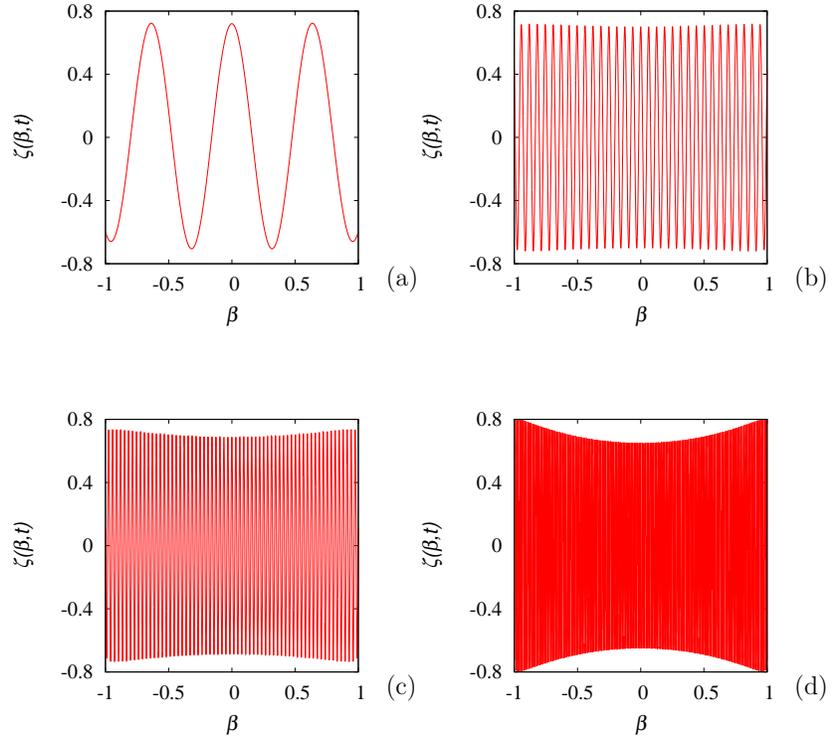}
\end{center}
\caption{ $\zeta(\beta,t)$ vs $\beta$ at different time instants for
$k^*=10$ obtained by solving eq. (\ref{eq3_11}) starting from a uniform initial
profile. (\textbf{a}) refers to $t=1$, (\textbf{b}) to $t=10$, (\textbf{c}) to 
$t=20$,
(\textbf{d}) to $t=50$.}
\label{Fig9}
\end{figure}

\begin{figure}
\begin{center}
\includegraphics[width=8cm]{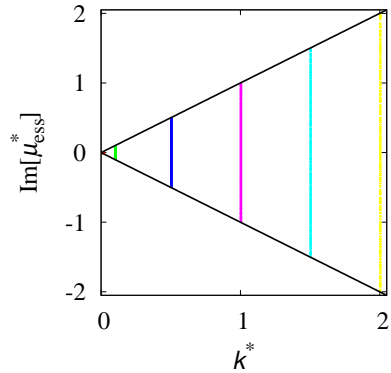}
\end{center}
\caption{Imaginary part of the essential eigenvalues 
$\mbox{Im}[\mu_{\rm ess}^*]$
vs $k^*$. Horizontal dots correspond to the calculated eigenvalues
at $k^*=0.01,\,0.1,\,0.5,\,1,\,1.5,\,2$. Solid lines represent
the lines $\mbox{Im}[\mu_{\rm ess}^*] = \pm k^*$.}
\label{Fig10}
\end{figure}

Consequently, in the statistical evolution of the process, the point spectrum plays
a marginal  role, especially for  high wavevectors, as
the dynamics is controlled by the essential  spectral component.
From eq. (\ref{eq3_14}) it follows that also
the essential spectrum is lower bounded, and this fact is
expressed by the property $\gamma^*=1$.
The typical evolution of a high-wavevector excitation,
above the critical value $k_c$, is depicted in figure \ref{Fig9},
showing the progressive development of a complex structure in $\beta$,
not converging to any proper eigenfunction.
By expanding $\zeta(\beta,t)$ in truncated  Fourier series
with respect to $\beta$, $\zeta(\beta,t)=\sum_{n=-N}^N\zeta_n(t) \, e^{i \, n \, \pi \beta}$, keeping $N$ sufficiently large,
the eigenvalue spectrum can be calculated by solving
the eigenvalue problem  $\sum_{k=-N}^N M_{n,k} \, \zeta_k= \mu^* \, \zeta_n$,
where $M_{0,0}=0$, $M_{n,k}= - k^* \, d(k-n)/2- \delta_{n,k}$, for $|n|+|k|>0$,
where $d(k)=2 (-1)^k/(k \, \pi)$, $|k|>0$.
The direct numerical calculation of the spectrum, obtained numerically
by setting $N=5000$, provides 
for $k^*<k_c$, only a single eigenvalue in the point spectrum, as already found,
while for all the values of $k^*$, the essential spectrum is
made by essential eigenvalues  $\mu_{\rm ess}^*$ possessing real part 
identically equal to
$-1$, and imaginary part  distributed in an uniform way within the
interval $[-k^*,k^*]$. This phenomenon is depicted in figure \ref{Fig10}.

\begin{figure}
\begin{center}
\includegraphics[width=8cm]{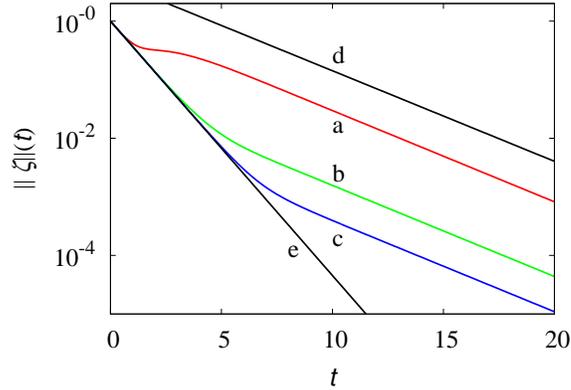}
\end{center}
\caption{$||\zeta||(t)$ vs $t$  at $k^*=1$ for several initial
conditions, $\zeta(\beta,0)=\sin(\nu \, \pi \, \beta)$.
. Lines (a) to (c) refer To $\nu=1,\, 10,\,40$,
respectively.  Line (d) represents the decay $||\zeta||(t) \sim e^{-r \, t}$
controlled by the point-spectrum eigenvalue $r \simeq 0.356$,
line (e) the decay $||\zeta(t)||(t)\sim e^{-t}$ associated with the essential
spectrum.} 
\label{Fig11}
\end{figure}

The overall dynamics of the process involves both the spatial
evolution (i.e. the dynamics with respect to the variable $x$), parametrized
by the wavevector $k$ (upon a Fourier transform), and the internal
dynamics with respect to the variable $\beta$, describing the
recombination process amongst the partial density waves.
Below the critical threshold $k_c$, the unique point spectrum eigenvalue
controls the long-term exponential relaxation of the solutions of
eq. (\ref{eq3_11}). However, the initial/intermediate decay is
sensitive to the relaxation exponent $\gamma^*$ pertaining to the
essential spectrum. This initial/intermediate behavior becomes
more evident starting from initial conditions $\zeta(\beta,0)$
possessing high-frequency components with respect to the internal
variable $\beta$. This phenomenon can be clearly
observed from the data in figure \ref{Fig11} showing the decay of
the $L^2$-norm $||\zeta||(t)$ of $\zeta(\beta,t)$ at $k^*=1$,
starting
from differential initial sinusoidal profiles,
$\zeta(\beta,0)= \sin(\nu \, \pi \, \beta)$, $\nu$ integer.
For $\nu \sim O(1)$, only the long-term exponential
relaxation can be observed (line a), the exponent of which is
the single point-spectrum exponent $r \simeq 0.356$. 
As $\nu$ increases, $\nu >10^1$,
a crossover between the initial/itermediate exponential scaling
depending on the essential component of the spectrum $||\zeta||(t) \sim
e^{-\gamma^* \, t}$, and the long-term relaxation
 $||\zeta ||(t) \sim e^{-r t}$ occurs.

As a final observation, let us consider the decay of the impulsive
component of the overall density function
for  an initial preparation
of the system with particles possessing  one and the 
same initial velocity, i.e.,
$p(x,t;\beta)=\delta(x) \, \delta(\beta-\beta^*)$.
This situation corresponds  qualitatively to the evolution of the density profiles depicted in figure \ref{Fig2} panel (a) for the
simpler one-velocity model.
in this case the overall density
$P(x,t)$ at time $t$ is the superposition of
a continuous distribution $P_c(x,t)$ and of an impulsive component
centered at $x= b_0 \, \beta^* \, t$,
\begin{equation}
P(x,t;)= P_c(x,t)+ I_{\beta^*}(t) \, \delta(x-b_0 \, \beta^* \, t) \, .
\label{eq3_15}
\end{equation}
The impulsive component $I_{\beta^*}(t) $ translating in  space at
velocity  $\beta^*$, corresponds to the fraction of particles
that remain in the initial state $\beta^*$, i.e. 
 not experiencing any transition up to time $t$. Because of its
physical meaning, it is clear that the relaxation of $I_{\beta^*}(t)$
should not depend on the spatial component of the dynamics
of $p(x,t;\beta)$, i.e., should be independent of the wavevector
$k$. Not only this, from the above observation it follows
that it is a property exclusively of the Markov recombination mechanism
of the partial density waves, described by the internal state dynamics
eq. (\ref{eqx_2}), and associated with the internal state operator
$\widehat{\mathcal L}$
\begin{equation}
\widehat{\mathcal L}[f(\beta)]= -\lambda_0 \, f(\beta) +
\frac{\lambda_0}{2} \int_{-1}^1 f(\beta^\prime) \, d \beta^\prime \, .
\label{eq3_16}
\end{equation}
The spectrum of $\widehat{\mathcal L}$ is completely degenerate:
it possess only the single eigenvalue $\mu=\lambda_0$ with
countable multiplicity, as any function $\pi(\beta)$ possessing zero
mean in $[-1,1]$ is an eigenfunction associated to it.
Specifically, a basis for this eigenspace is given by the
sinusoidal functions $\pi_\nu(\beta)=\sin(\nu \, \pi \, \beta)$,
where $\nu=1,\,2,\,...$.
This eigenproperty of the recombination mechanism, when embedded
into the spatio-temporal evolution of the partial
density waves $p(x,t,\beta)$ defined by eq. (\ref{eqx_5}) results
incompatible with the translation operator $-b_0 \, \beta \, \nabla$ 
in order to determine any proper eigenstructure, and thus
contribute to the essential part of the spectrum of the
operator ${\mathcal L}_\beta$ associated with the spatial-temporal
evolution of the partial density waves.
 
\begin{figure}
\begin{center}
\includegraphics[width=8cm]{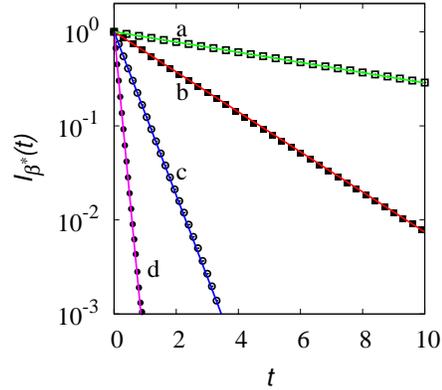}
\end{center}
\caption{Intensity of the impulsive contribution $I_{\beta^*}(t)$ vs $t$
obtained from stochastic simulations (symbols)  at $\beta^*=1/2$,
for different values of $b_0$,
keeping constant the ratio $b_0^2/2 \lambda_0=1$. The solid lines represent
the scaling $I_{\beta^*}(t)=e^{- \lambda_0 \, t}$.
Line (a) and  symbols  ($\square$)  refer to $b_0=0.5$,
line (b) and symbols ($\blacksquare$) to $\beta^*=1$,
line (c) and symbols ($\circ$) to $\beta^*=2$,
line (d) and symbols ($\bullet$) to $\beta^*=4$.}
\label{Fig12}
\end{figure}

It follows from the above discussion, that the impulsive part of
the dynamics $I_{\beta^*}(t)$ entering eq. (\ref{eq3_16})
should decay exponentially as
\begin{equation}
I_{\beta^*}(t) \sim e ^{-\lambda_0 \, t} \, .
\label{eq3_17}
\end{equation}
corresponding to the decay defined by the essential spectral component.
This phenomenon is depicted in figure \ref{Fig12}. The numerical
data have been obtained from the stochastic simulation of an
ensemble of $N_p=10^8$ particles evolving in space according to
eq. (\ref{eqx_5}), initially placed at $x=0$ and all possessing
the same velocity $b_0 \, \beta^*$. Therefore, this
preparation of the system corresponds to the initial
condition for the partial density waves
expressed by $p(x,0;\beta)= \delta(x) \, \delta(\beta-\beta^*)$.
By tracking the evolution of the particle system, the intensity
of the impulsive peak $I_{\beta^*}(t)$ can be determined.

Data in figure \ref{Fig12} refer to different values
of $b_0$, keeping fixed to ratio $b_0^2/2 \lambda =1$, so
that the decay of the impulsive peak depends on the velocity  $b_0$ as
$I_{\beta^*}(t) \sim e^{-b_0^2 \, t/2}$.

\section{Quantum mechanical extension}
\label{sec6}

There is  an intertwined network of mutual analogies
between 
the theory of Brownian motion  (and more generally 
stochastic dynamics) and quantum
mechanics 
 that has led to fruitful physical insigths and useful
mathematical representations, borrowing ideas and techniques from one
field to the other \cite{nelson,nelson_book,gudder,naga}.
{Specifically, there is a  strong connection between stochastic processes
possesing finite propagation velocity and  quantum mechanics, whenever the  relativistic constraint
on the upper bound for velocities in the space-time propagation of physical phenomena is taken into
account. This was already observed by Feynman \cite{feynman} in his development of the path-integral formulation
of quantum mechanics, as regards the path of a relativistic free particle (see Fig.~2.4 in \cite{feynman}).
Hovewer, Gaveau et al. \cite{gaveau} provided the derivation of the strict analogy between the $1+1$ Dirac
equation (1 spatial dimension + 1 temporal dimension)  and Poisson-Kac processes in the presence of
imaginary transition rates,  i.e.,
\begin{eqnarray}
\frac{\partial \psi_+(x,t)}{\partial t} & = & - c \, \frac{\partial \psi_+(x,t)}{\partial t} + i \, \lambda_0 \,
\left [ \psi_+(x,t)-\psi_-(x,t) \right ] \nonumber \\
\frac{\partial \psi_-(x,t)}{\partial t} & = &  c \, \frac{\partial \psi_-(x,t)}{\partial t} + i \, \lambda_0 \,
\left [ \psi_-(x,t)-\psi_+(x,t) \right ] \, ,
\label{eqadd}
\end{eqnarray}
where $c$ is the velocity of light {\em in vacuo}, $i= \sqrt{-1}$ and $(\psi_+,\psi_-)$ are the
two components of the vector-valued wave function. The  spatial probability density is
$|\psi_+|^2+|\psi_-|^2$, consistent with the Dirac theory \cite{diracth}.
As for the probabilistic theory of Poisson-Kac processes, eq.~(\ref{eqadd}) converges towards the
Schr\"odinger equation in the limit of $c,\lambda_0 \rightarrow \infty$, keeping fixed the ratio
$c^2/2 \lambda_0$ (Kac limit). This model has been analyzed and generalized  by several authors
\cite{jona,ord,birula,strauch,bala}.}

{In this Section we extend the stochastic model
considered in the previous paragraphs to a quantum mechanical
setting following the approach by Gaveau et al. \cite{gaveau} discussed above, by considering a continuous velocity instead of a single velocity $c$, as in  (\ref{eqadd}).}
  The reason for this extension, stems from the
observation that the complete description of the dynamics
of the stochastic process (\ref{eqx_5}) requires  the
introduction of the addition variable $\beta$  parametrizing
the partial density functions $p(x,t;\beta)$.
In a quantum extension, this variable can be viewed
as a ``hidden variable'' associated with a
sub-quantum level of description of the system, from
which classical quantum theory can be viewed as an emergent
property. In other words, this quantum extension is aimed
at providing an archetype for possible sub-quantum
descriptions of reality, in the spirit of the work
by David Bohm \cite{bohm1,bohm2}, showing how from this ``hidden'' level,
classical  quantum theory can be emergently derived.
As we are primarily interested in highlighting these two
aspects, the simple case of a free particle is considered
in  a one-dimensional spatial setting, in agreement with 
the stochastic model developed in the previous Sections.

The quantum mechanical 
counterpart of eq. (\ref{eqx_5}) is given by
\begin{equation}
\frac{\partial \psi(x,t;\beta)}{\partial t}= - b_0 \, \beta \, \frac{\partial
\psi(x,t;\beta)}{\partial x} + i \, \lambda_0 \, \left
[ \psi(x,t;\beta) - \frac{1}{2} \int_{-1}^1 \psi(x,t;\beta^\prime) \, d \beta^\prime \right ] \, ,
\label{eq4_1}
\end{equation}
where $\psi(x,t;\beta)$ is the wavefunction at the
subquantum level, parametrized with respect to the
additional variable $\beta \in [-1,1]$. 
The constants
$b_0$ and $\lambda_0$ are determined below.

Let $\Psi(x,t)$ be the classical (Schr\"odinger) wavefunction,
\begin{equation}
\Psi(x,t)= \int_{-1}^1 \psi(x,t;\beta) \, d \beta \, ,
\label{eq4_2}
\end{equation}
and introduce the flux $J_\psi(x,t)= b_0 \, \int_{-1}^1 \beta
\, \psi(x,t;\beta) \, d \beta$, such that
\begin{equation}
\frac{\partial \Psi(x,t)}{\partial t}= - \frac{\partial J_\psi(x,t)}{\partial
x} \, .
\label{eq4_3}
\end{equation}
From eq. (\ref{eq4_1}) it follows that $J_\psi(x,t)$ fulfils the
equation
\begin{equation}
\frac{\partial J_\psi(x,t)}{\partial t}=
- b_0^2 \, \frac{\partial }{\partial x} \int_{-1}^1 \beta^2 \, \psi(x,t;\beta)
\, d \beta + i \, \lambda_0 \, J_\psi(x,t) \, .
\label{eq4_4}
\end{equation}
Let us consider the Kac limit of this model, letting $b_0, \, \lambda_0 \rightarrow
\infty$, keeping constant the nominal diffusivity $b_0^2/2 \lambda_0=D_{\rm nom}$. From eq. (\ref{eq4_1}), in the Kac limit, 
\begin{equation}
\psi(x,t;\beta)= \frac{1}{2} \int_{-1}^1 \psi(x,t;\beta) \, d \beta = \frac{\Psi(x,t)}{2} \, , 
\label{eq4_5}
\end{equation}
i.e. $\psi(x,t;\beta)$ in the Kac limit does not depend on $\beta$,
while  $J_\psi(x,t)$ attains the expression
\begin{equation}
J_\psi(x,t) = - i  \frac{b_0^2}{2 \, \lambda_0} \, 2 \frac{\partial }{
\partial x} \int_{-1}^1 \beta^2 \, \psi(x,t;\beta) \, d \beta
= - i \frac{2 \, D_{\rm nom}}{3} \, \frac{\partial \Psi(x,t)}{\partial x} \, ,
\label{eq4_6}
\end{equation}
where eq. (\ref{eq4_5}) has been used. From eq. (\ref{eq4_6}) inserted in
eq. (\ref{eq4_3}) it follows that if $D_{\rm nom}$ is such that
$D_{\rm nom}= 3 \, D_{\hbar}/2$, where $D_{\hbar}= \hbar/2 \, m$
is the quantum diffusivity for a massive particle of mass $m$, then
$\Psi(x,t)$ satisfies the Schr\"odinger equation
\begin{equation}
- i \, \frac{\partial \Psi(x,t)}{\partial t}= D_{\hbar} \, \frac{\partial^2 \psi(x,t)}{\partial x^2} \, .
\label{eq4_7}
\end{equation}
Moreover, it is easy  to show that eq. (\ref{eq4_1}) is a proper quantum mechanical
equation, in the meaning that a Born-rule can be defined from it.
In point of fact, elementary calculations provides
\begin{equation}
\frac{\partial }{\partial t} \int_{-1}^1 |\psi(x,t;\beta)|^2 \, d \beta
=
- \frac{\partial }{\partial x} \left (
b_0 \, \int_{-1}^1 \beta |\psi(x,t;\beta) |^2 \, d \beta \right ) \, .
\label{eq4_8}
\end{equation}
and therefore $\int_{-1}^1 |\psi(x,t;\beta)|^2 \, d \beta$ is
a conserved quantity corresponding to the quantum probability density
function with respect to the position $x$. {Observe that, in the Kac limit,
$\psi(x,t;\beta) \sim \Psi(x,t)$, and  therefore $\int_{-1}^1 |\psi(x,t;\beta)|^2 \, d \beta \sim |\Psi(x,t)|^2$,
in agreement with the Schr\"odinger's nonrelativistic quantum theory.}

{For the sake of completeness, it is useful to observed that eq. (\ref{eq4_1}) refers to
a  single free particle, but its extension to include the action of a potential or its generalization to
a  many-body problem
is straightforward. If $U(x)$ is a potential acting on a massive particle,  the associated quantum
model generalizing eq. (\ref{eq4_1}) is expressed by
\begin{equation}
\frac{\partial \psi(x,t;\beta)}{\partial t}= - b_0 \, \beta \, \frac{\partial
\psi(x,t;\beta)}{\partial x} - i \,  \frac{U(x)}{\hbar} \, \psi(x,t;\beta)+ i \, \lambda_0 \, \left
[ \psi(x,t;\beta) - \frac{1}{2} \int_{-1}^1 \psi(x,t;\beta^\prime) \, d \beta^\prime \right ] \,  .
\label{eq4_b1}
\end{equation}
Using the same approach outlined above, it is easy to see that eq. (\ref{eq4_1}) converges in the Kac limit
to the nonrelativistic Schr\"odinger equation in the presence of the potential $U(x)$.
In a similar way, the extension to a many-body problem is formally  simple.  To begin with,
consider a single spatial dimension.
 Let ${\bf x}=(x_1,...,x_n)$ be the coordinate vector of the particle system,
where $x_k$ is the position of the $k$-th particle,  $k=1,\dots,n$, and $\boldsymbol{\beta}=(\beta_1,\dots,\beta_n)$,
$\boldsymbol{\lambda}=(\lambda_1,\dots,\lambda_n)$, so that $b_0 \, \beta_k$ is the velocity of the
$k$-th particle with $\beta_k \in [-1,1]$, and $\lambda_k$ is its transition rate.
The extension of eq. (\ref{eq4_1}) to  an $n$-body problem on the real line in the presence
of the interaction potential $U({\bf x})$ reads,
\begin{eqnarray}
\frac{\partial \psi({\bf x},t;\boldsymbol{\beta})}{\partial t} & = & - b_0 \, \sum_{k=1}^n \beta_k \,
\frac{\partial \psi({\bf x},t;\boldsymbol{\beta})}{\partial x_k} - i \, \frac{U({\bf x})}{\hbar}  \psi({\bf x},t;\boldsymbol{\beta}) \nonumber \\
& + & i \, \sum_{k=1}^n \lambda_k \, \left [ \psi({\bf x},t;\boldsymbol{\beta}) - \frac{1}{2^n} \int_{-1}^1  \cdots
\int_{-1}^1 \psi({\bf x},t;\boldsymbol{\beta}^\prime) \, d \boldsymbol{\beta}^\prime \right ] \, ,
\label{eq4_b2}
\end{eqnarray}
where $d \boldsymbol{\beta}^\prime=d \beta_1^\prime \cdot \dots \cdot d \beta_n^\prime$.  
In higher dimensions, considering only translational motion, let us define  ${\bf X}=({\bf x}_1,\dots,{\bf x}_n)$,
where ${\bf x}_k=(x_{k,1},\dots,x_{k,d})$ is the position vector of the $k$-th particle in a $d$-dimensional
space, and
$\overline{\boldsymbol{\beta}}=(\boldsymbol{\beta}_1,\dots,\boldsymbol{\beta}_n)$, where the velocity  of the
$k$-th particle is $b_0 \, \boldsymbol{\beta}_k= b_0 (\beta_{k,1},\dots,\beta_{k,d})$,  and assuming isotropic
transition rates, eq. (\ref{eq4_b2}) simply becomes
\begin{eqnarray}
\frac{\partial \psi({\bf X},t;\overline{\boldsymbol{\beta}})}{\partial t} & = & - b_0 \, \sum_{k=1}^n \boldsymbol{\beta}_k  \cdot
\nabla_k \psi({\bf X},t;\overline{\boldsymbol{\beta}}) - i \, \frac{U({\bf X})}{\hbar}   \, \psi({\bf X},t;\overline{\boldsymbol{\beta}})
\nonumber \\
& + & i \, \sum_{k=1}^n  d \, \lambda_k \, \left [ \psi({\bf X},t;\overline{\boldsymbol{\beta}}) - \frac{1}{2^{n \, d}} \int_{-1}^1  \cdots
\int_{-1}^1 \psi({\bf X},t;\overline{\boldsymbol{\beta}}^\prime) \, d \overline{\boldsymbol{\beta}}^\prime \right ] \, ,
\label{eq4_b3}
\end{eqnarray}
where $\nabla_k= \left ( \partial /\partial x_{k,1}\dots \partial /\partial x_{k,d} \right )$, and
$d \overline{\boldsymbol{\beta}}^\prime=\prod_{k=1}^n \prod_{h=1}^d d \beta_{k,h}^\prime$.
The inclusion of $n$-body interactions
expressed by the potential $U({\bf X})$ is perfectly admissible in a non-relativistic framework, while
it would require care and attention in a relativistic perspective. But this discussion is manifestly outside
the scope of the present work.}

\subsection{Spectral properties and dispersion relations}
\label{sec4_1}

\begin{equation}
{\mathcal L}_q[\phi(x,\beta)] = - b_0 \, \beta \frac{\partial \psi(x,\beta)}{\partial x} + i \, \lambda_0 \left [ \psi(x,\beta) - \frac{1}{2} \int_{-1}^1 \psi(x,\beta^\prime) \, d \beta^\prime \right ] \,, 
\label{eq4_9}
\end{equation}
and let $\varepsilon$ be  an eigenvalue, $\mathcal L_q[\phi(x,\beta)]=\varepsilon \, 
\phi(x,\beta)$. The eigenfunctions are of the form
$\phi(x,\beta)= e^{i \, k \, x} \, \psi_0(\beta)$.
Introducing, as before,  the normalized quantities, $\varepsilon^*=\varepsilon/\lambda_0$, $k^*= k \, b_0/\lambda_0$, it follows that
that the ``internal component'' of the eigenfunction $\phi_0(\beta)$
associated with the normalized eigenvalue $\varepsilon^*$ is given
by
\begin{equation}
\phi_0(\beta)= \frac{A}{\varepsilon^*-1 + k^* \, \beta} \, ,
\label{eq4_10}
\end{equation}
where $A$ is a constant, leading to the eigenvalue equation
\begin{equation}
-2 = \int_{-1}^1 \frac{d \beta}{\varepsilon^* -1 + k^* \, \beta} \, ,
\label{eq4_11}
\end{equation}
that provides, upon quadraturae and algebraic manupulations,
the expression for the eigenvalue
\begin{equation}
\varepsilon^* = 1 - \frac{k^* \, (1+ e^{-2 \, k^*})}{1-e^{-2 \, k^*}} \, .
\label{eq4_12}
\end{equation}
In the quantum case, for any wavevector $k$ there exists
an eigenvalue in the point spectrum of the operator ${\mathcal L}_q$,
which is not the case for its probabilistic counterpart ${\mathcal L}_\beta$
as thoroughly analyzed in Sections \ref{sec4}-\ref{sec5}.

Let us consider the scaling properties of the eigenvalue spectrum,
i.e., the dispersion relation connecting the particle
energy $E= - \hbar \, \varepsilon$ to the wave vector $k$.

The Kac limit ($b_0, \, \lambda_0  \rightarrow \infty$, $b_0^2/2 \, \lambda_0=\mbox{constant}$), corresponds to small values
of $k^*$. Expanding the exponential entering
 eq. (\ref{eq4_12}) in a Taylor series 
to the
leading order, in this case $e^{-2 \, x}=1-2 x+2 x^2 -4 x^3/3$,
and developing the necessary algebra, one obtains from eq. (\ref{eq4_12})
\begin{equation}
\varepsilon^* = - \frac{(k^*)^2}{3} + O((k^*)^3) \, .
\label{eq4_13}
\end{equation}
Neglecting the $O((k^*)^3)$ term, it implies for the eigenvalue $\varepsilon$,
\begin{equation}
\varepsilon = \lambda_0 \, \varepsilon^* = - \frac{\lambda_0}{3} \, \frac{b_0^2 \, k^2}{\lambda_0^2} = - \frac{2 \, D_{\rm nom}}{3} \, k^2 = - D_\hbar \, k^2 \, ,
\label{eq4_14}
\end{equation}
and consequently the energy is given by
\begin{equation}
E = - \hbar \, D_\hbar \, k^2 = \frac{\hbar^2 \, k^2}{2 \, m} \, .
\label{eq4_15}
\end{equation}
consistently with the classical quantum mechanical result.

\begin{figure}
\begin{center}
\includegraphics[width=8cm]{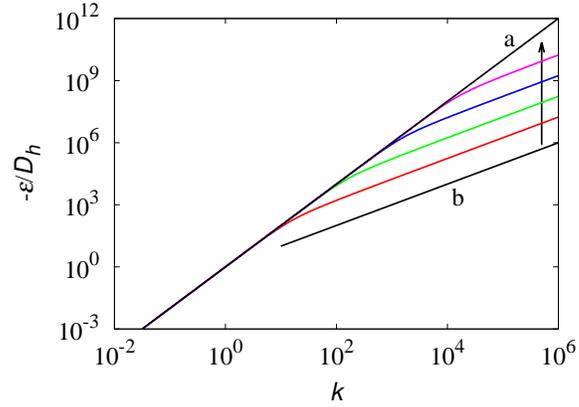}
\end{center}
\caption{Dispersion curve $-\varepsilon/D_h$ vs the wavenumber $|k|$
for a free particle in the sub-quantum model (\ref{eq4_1}).
Solid line (a) is the classical quantum result $-\varepsilon/D_h= k^2$.
Solid line (b) represents the linear scaling $-\varepsilon/D_h \sim |k|$.
The other lines depict the graph of the function at the r.h.s of eq. (\ref{eq4_16})
for increasing values  of $\sigma=10^2,\,10^4,\,10^6,\,10^8$, indicated
by the direction of the arrow.}
\label{Fig13}
\end{figure}

Consider the quantity $-\varepsilon/D_\hbar$, that in the
classical quantum limit equals $k^2$, and set $\sigma= \lambda_0/D_\hbar$.
Since $k^*= \sqrt{3/\sigma} \, k$ is follows that
\begin{equation}
-\frac{\varepsilon}{D_\hbar}= -\sigma
\, \left [ \frac{(1-e^{2 \, k^*}) - k^* \, (1+e^{-2 \, k^*})}{1-e^{-2 \, k^*}}
\right ]_{k^*= \sqrt{3/\sigma} \, k} \, .
\label{eq4_16}
\end{equation}
For large values $|k|$, $-\varepsilon/D_\hbar= \sigma |k^*|$, and
therefore $-\varepsilon/D_\hbar$ admits the crossover
behavior
\begin{equation}
-\frac{\varepsilon}{D_\hbar} =
\left \{
\begin{array}{ccc}
k^2  \, ,& & |k| < \sqrt{3 \,\sigma} \\
\sqrt{3 \, \sigma} \, |k| \, ,  & & |k| > \sqrt{3 \,\sigma}
\end{array}
\right . \, .
\label{eq4_17}
\end{equation}
This phenomenon is depicted in figure \ref{Fig13}.
The linear scaling with $|k|$ in the high wavevector region is
the signature in the  quantum model of the bounded
propagation velocity characterizing the generalized
Schr\"odinger operator ${\mathcal L}_q$ for
finite values of $b_0$ and $\lambda_0$.

\section{Concluding remarks}

In this article we have analyzed
the spectral (eigenvalue) properties of several classes
of stochastic processes possessing finite propagation velocity,
showing the occurrence of a lower bound for the real part of the
eigenvalue spectrum.
 As regard this
spectral property, it is immaterial whether the long-term diffusive
behavior is regular (linear Einsteinian scaling of the mean square displacement)
or anomalous (superdiffusive).
This result marks a fundamental and significant difference with
respect to the parabolic diffusion model, for which
  short wavelength perturbations
decay at a rate proportional to the squared norm of the wavevector.

There are several general implications of this result. (i) In the
approach to the foundations of the thermodynamics of irreversible
processes \cite{mackey}: The results obtained confirm the analysis
developed in \cite{giona_physa} indicating that the Markov operators
associated with SEOs of stochastic processes characterized by a finite
propagation velocity are invertible and form a group of
transformations parametrized with respect to time $t$.  Conversely, in
the case of parabolic diffusion models the weaker semigroup property
for $t \geq0$ holds. (ii) In classical field and transport theory: The
results indicate that the hyperbolic field equations (e.g. for mass,
heat and momentum transfer) that can be developed starting from
microscopic equations of motions expresses in the form of stochastic
equations, characterized by a bounded velocity of propagation
\cite{gpk3}, may have completely different stability properties than
their parabolic counterparts.

The occurrence of a lower bound for the real part of the eigenspectrum
 of  operators describing the statistical  evolution
of stochastic processes possessing finite propagation velocity
occurs also for continuum models with a spectrum of particle
velocities. In this article we have considered a prototypical
example on the real line, in which the relaxational properties
at high wavevectors are entirely associated with
the essential part of the spectrum. The lower spectral bound
is  a consequence of  the internal recombination dynamics of the
stochastic states of the system, and corresponds to the
largest decay  exponent (with reversed sign) of the overall
probability density function $P(x,t)$, for an initial configuration
with all the particles located at the same point and possessing the same 
velocity.

{In a broader perspective, the analysis of SEOs outlined
  in this article can be extended to other relevant physical
  phenomenologies.  This is the case for the linear response of
  stochastic systems possessing anomalous behavior, addressed in
  \cite{lr1,lr2,lr3,lr4,lr5} by either using Continuous Time Random
  Walk scalings or approximate SEOs involving fractional derivative
  operators. The same problem can by framed within the age formalism
  of LWs introduced in Section \ref{sec3}, by keeping the same
  definitions and evolution equations (\ref{eq6_1})-(\ref{eq6_3}), and
  modifying the boundary condition eq. (\ref{eq6_4}) in the form of
\begin{equation}
p_\pm(x,t;0)= A_{\pm,+}(t) \, \int_0^\infty \lambda(\tau) \,
p_+(x,t;\tau) \, d \tau + A_{\pm,-}(t) \, \int_0^\infty \lambda(\tau) \,
p_-(x,t;\tau) \, d \tau\:,
\label{eqx_c1}
\end{equation}
where the transition probabilities $A_{\pm,\pm}(t)$, $A_{\pm,\pm} (t)
\geq 0$, $A_{+,\pm}(t)+A_{-,\pm}(t)=1$ account for the effect of the
forcing field $f(t)$ and are defined according to \cite{lr1,lr4} as
\begin{equation}
A_{+,\pm}(t) = \frac{1+ \mu \, f(t)}{2} \, , \qquad
A_{-,\pm}(t) = \frac{1- \mu \, f(t)}{2}\:,
\label{eqx_c2}
\end{equation}
where $0< \mu < 1$ and $|f(t)|\leq 1$. In the analysis of this problem
the spectral results outlined in this manuscript are of limited use, as
the simplest way to tackle this problem from physical grounds is to
consider the partial moment hierarchy $m_\pm^{(n)}=\int_{-\infty}^\infty
x^n \,  p_\pm(x,t;\tau) \, d x$, which will be addressed in forthcoming works.}

Getting back to our probabilistic model with a spectrum of velocities,
its quantum extension displays
two main qualitative properties: (i) contrarily to the
purely probabilistic case, for any $k \in {\mathbb R}$ an eigenvalue of the
quantum operator ${\mathcal L}_q$ exists; (ii) the boundedness in the
propagation velocity characterizing ${\mathcal L}_q$ implies
in the quantum case a linear dispersive relation of the
energy $E$ with respect to $|k|$ for high values of $k$. This result is a direct consequence
of the undulatory propagation, as from quantization $E = \hbar \, \omega$,
and the $\omega$ is related to the group velocity $v_g$, by 
$d \omega/d k=v_g$, implying $E \sim k$.
The quantum model analyzed in paragraph \ref{sec4} is a simple
example of 
a sub-quantum model possessing
internal (''hidden'') degrees of freedom (in this
case the internal variable $\beta \in [-1,1]$) and providing
the same emergent results of classical quantum theory.

\funding{This research received no external funding.}

\conflictsofinterest{The authors declare no conflict of interest.}

\abbreviations{Abbreviations}{
The following abbreviations are used in this manuscript:\\

\noindent
\begin{tabular}{@{}ll}
SEO & Statistical evolution operator \\
GPK & Generalized Poisson-Kac \\
LW & L\'evy Walk\\
\end{tabular}}

\end{paracol}
\reftitle{References}

\end{document}